\documentclass[prl,aps,epsf,twocolumn,floats,showpacspsfig,amsmath]{revtex4}
\usepackage{amssymb}

\usepackage{epsfig}

\newcommand{\be}{\begin{equation}}
\newcommand{\ee}{\end{equation}}
\newcommand{\bea}{\begin{eqnarray}}
\newcommand{\eea}{\end{eqnarray}}

\newcommand{\p}{\partial}

\newcommand{\la}{\langle}
\newcommand{\ra}{\rangle}
\newcommand{\rd}{\mbox{d}}
\newcommand{\ri}{\mbox{i}}
\newcommand{\re}{\mbox{e}}

\begin{document}
\title{Exactly solvable model for isospin S=3/2 
fermionic  atoms on an optical lattice}

\author{D. Controzzi$^a$,  and  A.  M. Tsvelik$^{b}$ }
\affiliation{ $^a$
International School for Advanced Studies and INFN, via Beirut 4,
34014 Trieste, Italy,\\ 
$^{b}$ Condensed Matter Physics and Materials Science Department, Brookhaven National Laboratory, Upton, NY 11973-5000, USA}
\date{\today}

\begin{abstract}
We propose an exact solution of a model 
describing a low energy
behavior of cold isospin S=3/2 fermionic atoms on a one-dimensional  optical
lattice. Depending on the band filling the effective field theory has
a form of a deformed Gross-Neveu model with either  
$O(7)\times \mathbb{Z}_2$ (half filling) or $U(1)\times O(5)\times
\mathbb{Z}_2$ symmetry.  
\end{abstract}

\pacs{PACS numbers: 71.10.Pm, 72.80.Sk}
\maketitle

\sloppy
 
Although high symmetries do not occur frequently in nature, they deserve
attention since  every new
symmetry brings   with itself a possibility of new physics. According to \cite{zhang}, the SO(5)
and SO(7) symmetries can be realized without fine-tuning
of parameters in lattice models of isospin-3/2 fermions  with a contact interaction. An experimental  realization was found in \cite{PhD}, where a four component gas of fermionic  K$^{40}$ atoms was kept in an optical trap for 50 seconds.  
 In this letter we will limit ourselves to one dimensions where the corresponding models happen to be
exactly solvable. The appropriate model is the one-band generalized Hubbard model \cite{zhang,ho}
\bea
 H &=& -t\sum_{i,\alpha}[\psi^+_{\alpha}(i+1)\psi_{\alpha}(i) +
  h.c.] - 
\mu\sum_{i,\alpha}\psi^+_{\alpha}(i)\psi_{\alpha}(i) \nonumber\\
 && + U_0\sum_i P_{0,0}^+(i)P_{0,0}(i) + U_2\sum_{i,m}P_{2,m}^+(i)P_{2,m}(i)
 \label{model}
\eea
where $\alpha$ and $m$ take values $\alpha = \pm 3/2,\pm 1/2$ and 
$m=0,\pm 1,\pm 2$. The singlet (total spin $S_T=0$) and quintet ( $S_T=2$)
pairing operators $P_{0,0}^+,P_{2,m}^+$ are defined through the
Clebsch-Gordon coefficients 
\[
P^+_{F,m}(i) = \sum_{\alpha,\beta}\la
\frac{3}{2}\frac{3}{2};F,m|\frac{3}{2}\frac{3}{2};\alpha\beta\ra\psi^+_{\alpha}(i)\psi^+_{\beta}(i),  
\] 
where $F=0,2$ and $m=-F,-F+1,...F$. In \cite{zhang} it was shown that
this model, together  with its continuous version,  have  a precise
U(1)$\times$SO(5) symmetry at arbitrary filling and  SO(7) symmetry at
half filling. 

It was demonstrated later that, at least in the low
energy limit, the total symmetry of these models is greater being
U(1)$\times$O(5)$\times\mathbb{Z}_2$ and SO(7)$\times\mathbb{Z}_2$ respectively
\cite{wu,azaria}.  Following the standard  procedure (see, for example, \cite{book}), 
at temperatures much
smaller than the Fermi energy one can linearize the
fermionic spectrum in the vicinity of the right and left Fermi points and to arrive to  the effective field theory description. If the band filling is incommensurate, the charge sector
decouples from the  other degrees of freedom at low energies. In this
limit the Hamiltonian can be decomposed into a sum of the charge and
the spin part $H \approx H_{charge} + H_{spin}$, where the charge
sector is described by the Gaussian model. All nontrivial physics is
concentrated in the spin Hamiltonian. It has SO(5)$\times\mathbb{Z}_2$
symmetry. The four-fermion interaction can be written as a sum of
products  of the SU$_1$(4) currents of the right and left chirality
\cite{azaria}, though its symmetry is lower than SU(4). To make the
true symmetry manifest one can employ the nonlinear transformation of
fields suggested in \cite{maldacena} and rewrite the Hamiltonian
density in the spin sector in terms of six species of Majorana
fermions (the bosonized form of this Hamiltonian was obtained in \cite{wu}):  
\bea 
&& {\cal H} = - \frac{\ri
}{2}\sum_{a=0}^5v_s(\bar \chi_a\p_x\bar \chi_a - \chi_a\p_x\chi_a) + 
\nonumber\\
&& g\left(\sum_{a=1}^5\bar\chi_a\chi_a\right)^2 +
2g_X(\bar\chi_0\chi_0)\left(\sum_{a=1}^5\bar\chi_a\chi_a\right).
\label{5+1} 
\eea
where $g \sim -(U_0 + U_2), g_X \sim -3U_2 + U_0$. One can run few
simple checks to justify that above model does represent  the
continuum limit of (\ref{model}).  First, the interaction in
(\ref{5+1}) has the right scaling dimension (in other words, this is
a model with four-fermion interaction). Second, at temperatures much
higher than the spectral gap but much smaller than the bandwidth, when
the interactions can be neglected, the two  models have the same
specific heat. This corresponds to model
(\ref{5+1}) having the right ultraviolet central charge $C_{UV} =
3$. Third, model (\ref{5+1}) has the symmetry predicted in
\cite{azaria} (the $\mathbb{Z}_2$ symmetry corresponds to $\chi_0
\rightarrow - \chi_0$). In the limit $g = g_X$ the
O(5)$\times\mathbb{Z}_2$ model  becomes the 
SO(6)$\equiv$SU(4) Gross-Neveu (GN) model. This limit is also contained
in original model (\ref{model}) \cite{azaria}. 
At half filling, due to the Umklapp processes, the charge mode is not
decoupled. Therefore the entire effective Hamiltonian is given by
Eq.(\ref{5+1}) with seven degenerate Majorana fermions instead of
five (the O(7)$\times\mathbb{Z}_2$ model).   

The O(5)$\times\mathbb{Z}_2$ (Eq.~\ref{5+1}) and  O(7)$\times\mathbb{Z}_2$ models (further down we refer to them as (5+1) and (7+1) models) have
nontrivial dynamics for $g > 0$, when they  scale to strong coupling
and spectral gaps are generated. As field theories they exist in two
limits, both corresponding to integrable models \cite{ContrTsv}. One limit,
associated to the largest symmetry (SO(6) or SO(8)), is realized
when $g = g_X$. The corresponding exactly  solvable model is the
SO(2N) ($N=3,4$) GN model \cite{ZamZamGN,ShankarWitten,Ann,OgReshW}.

In this letter we discuss the exact solution of the anisotropic
version of the models with $g \neq g_X$. Most of the calculations are conducted for the (5+1) case; at the end of the
paper we briefly discuss  the 
O(7)$\times\mathbb{Z}_2$ generalization. We construct the exact
solution using the bootstrap procedure as described, for instance, in
\cite{Ann}. Namely, guided by the symmetry of the model 
and the perturbation theory results we suggest the
two-particle S-matrix.  In integrable models, where 
multi-particle collisions are  representable as a sequence of independent two-particle ones (factorizability), such S-matrix contains all
information about the spectrum and the thermodynamics, as well as 
off-shell properties, such as correlation functions. The two-particle  S-matrix must satisfy the crossing,
unitarity and the Yang-Baxter conditions \cite{mcguire}. The latter one is a condition of associativity of the algebra of creation (annihilation) operators. This condition, being an overcomplete system of equations for the S-matrix elements, is very restrictive. Ones such solution is found, one has to complete the bootstrap process by calculating  the free energy and comparing it with  the perturbation theory for the model at hand. 

As we shall see, the bootstrap solution we obtain does not contain  
the fully symmetric (O(6) or O(8)) limit. This suggests 
that the symmetric  limit is unstable and cannot be reached asymptotically from
the exact solution of the asymmetric model. This conclusion is supported by the analysis  of the renormalization group (RG) equations. The RG equations for the O(2N+1)$\times\mathbb{Z}_2$
model are \cite{azaria1,wu,ContrTsv}
\bea
&&\dot g = - (2N-1)g^2 - g_X^2,\nonumber\\
&&\dot g_X = - 2Ng_Xg.
\eea 
They  have the following RG invariant: $C = (g^2 - g_X^2)/|g_X|^{b}, b =
(2N-1)/N$. From this it follows that the distance from  the
symmetric line $|g - g_X| \sim |C||g_X|^{(N-1)/N}$ does not decrease  
for all $N \geq 1$ under the increase in  $g_X$ rendering this line unstable. 
  
The integrability of the O(2N+1)$\times\mathbb{Z}_2$ model is a
natural extension of the $N=1$ case studied in detail in 
Refs.~\onlinecite{Ts87,andrei}.   In that case the  
two-particle S-matrix has the form 
\bea
S_{3+1} = \left(
\begin{array}{cc}
 [\hat S^{SU(2)}]_{\alpha,\beta}^{\bar\alpha,\bar\beta} & \zeta
 \delta_{\alpha}^{\bar\alpha} \\ 
\zeta \delta_{\alpha}^{\bar\alpha} & - 1
\end{array}
\right),
\eea
where $\alpha,\bar\alpha$ are $in$ and $out$ spinon indices,
$\zeta(\theta) = (\re^{\theta} - \ri)/(\re^{\theta} + \ri)$
and $\hat S^{SU(2)}$ is the S-matrix of SU(2) Thirring
model \cite{BKWK,Ann}. An important check of the validity of this
S-matrix is that the Thermodynamic Bethe ansatz equations constructed
from it yield the correct UV central charge $c_{3+1}=2$ and 
particle multiplicity.    
 
The S-matrix for the model (\ref{5+1}) was outlined  in
\cite{ContrTsv} as a generalization of the (3+1) case. In this letter we give further details of this solution. The suggested S-matrix has the following form:
\bea
S_{5+1} = \left(
\begin{array}{ccc}
\hat S_{vv} & \hat S_{vs} & -\hat I\\
\hat S_{vs} & \hat S_{ss} & \hat S_{s0}\\
-\hat I & \hat S_{s0} & - 1
\end{array}
\right).\label{S}
\eea 
The indices $s$ and $v$ label  massive kinks belonging to the 4-dimensional spinor representation and vector particles
(associated to the original fermions $\chi_a$ ( $a=1,...5$)). The kink's  mass is $M_s$; the vector particles have masses $M_v$. In the present
model they are not bound states of kinks. There is an
additional singlet particle (corresponding to the original $\chi_0$
fermion) with mass  $M_0$. All masses are expenentially small in the bare coupling constants. Since there are only two couplings in the theory, $M_s, M_v$ and $M_0$ are not independent, though to determine this dependence one needs a microscopic derivation.  
As one might have expected, the Majorana fermion $\chi_0$ acquires a
nontrivial phase factor by scattering on a kink:
\be
\hat S_{s0}(\theta)=\xi(\theta) \; \delta_{\alpha}^{\bar\alpha}~,~~~
\xi(\theta) = \frac{\re^{3\theta} - \ri}{\re^{3\theta} + \ri}
\ee
($\alpha,\bar\alpha$ are $in$ and $out$ kink indices) and has a
trivial S-matrix with the other vector particles. The rest of the
notations are as follows. $\hat S^{vv}$, $\hat S^{vs}$ and $\hat
S^{ss}$ are  O(5) S-matrices of vector 
and spinor particles. 
They can be written as  a
sum of projectors, $\hat P_a$, that map the tensor product of two 
representations onto the irreducible representation labeled $a$. For
instance, the spinor S-matrix has the form
\bea
\hat S_{ss}(\theta) = S_{ss}(\theta)\left[\hat P_{asym} + \frac{\theta +
    \ri\pi/3}{\theta - \ri\pi/3}\hat P_v + \frac{\theta +
    \ri\pi}{\theta - \ri\pi}\hat P_0\right], \label{spinor} 
\eea
where $\hat P_0, \hat P_v, \hat P_{asym}$ represent projectors onto
singlet, vector and antisymmetric tensor representations. The form of
the prefactors, $S_{ss},S_{sv},S_{vv}$, can be inferred from the
kernels (\ref{kernels}) via 
\bea
\label{S_kernels}
S_{ab}(\theta) = \exp\left(- \int\frac{\rd\omega}{\omega}\re^{-3\ri\theta\omega/\pi} Y_{ab}(\omega)\right),
\eea
and for the vector particles can be found in Ref.s~\onlinecite{ZamZamGN,Ann}.
For the spinor particles we have 
\bea
&& S_{ss}(\theta) = f(\theta)/f(-\theta), ~~ f(\theta) = \\
&& \prod_{k=0}^{\infty}\frac{\Gamma(2 + 3k + 3\ri\theta/2\pi)\Gamma(4 + 3k + 3\ri\theta/2\pi)}{\Gamma(5/2 + 3k + 3\ri\theta/2\pi)\Gamma(7/2 + 3k + 3\ri\theta/2\pi)}\nonumber
\eea
It has no poles on the physical sheet, thus, as we have already mentioned, spinors do not create bound states. 

The Bethe ansatz equations associated with  the S-matrix (\ref{S}) are
\begin{widetext}
\begin{subequations}
\label{BA1}
\bea
&& \re^{-\ri M_sL\sinh\theta_a^{(s)}} = \prod_{b=1}^{n_s}S_{ss}(\theta_a^{(s)}- \theta_b^{(s)})\prod_{b=1}^{n_v}S_{sv}(\theta_a^{(s)}- \phi_b^{(v)})\prod_{c=1}^{n_0}S_{s0}(\theta_a^{(s)}- u_c)\prod_{p=1}^{m_s}e_{1/2}(\theta_a^{(s)} - \lambda_p)\\
&&\re^{-\ri M_vL\sinh\phi_a^{(s)}} = \prod_{b=1}^{n_v}S_{vv}(\phi_a^{(v)}- \phi_b^{(v)})\prod_{b=1}^{n_s}S_{sv}(\phi_a^{(v)}- \theta_b^{(s)})\prod_{p=1}^{m_v}e_{1}(\phi_a^{(v)} - \mu_p)\\  
&& \re^{-\ri M_0L\sinh u_a} = (-)^{n_0+n_v}\prod_{b=1}^{n_s}S_{s0}(u_a -\theta_b^{(s)})
\eea
\end{subequations}
\end{widetext}
where the numbers $n_{s,v} > m_{s,v}$ and $n_0$ correspond to physical
particles (spinors, O(5) vector and singlet Majorana fermions respectively)
and  vanish in the ground state. 
The functions $e_n$ have the form $e_n(x) = (x + \ri n\pi/3)/(x - \ri n\pi/3)$ 
and the auxiliary particles satisfy
the eigenvalue equations for the O(5) transfer matrix \cite{OgReshW}
\bea
\prod_{b=1}^{m_v}e_1(\lambda_a -
\mu_b)\prod_{c=1}^{n_s}e_{1/2}(\lambda_a - \theta^{(s)}_c) =
\prod_{b=1}^{m_s}e_{1}(\lambda_a - \lambda_c)\nonumber \\
\prod_{b=1}^{m_s}e_1(\mu_a - \lambda_b)\prod_{b=1}^{n_v}e_1(\mu_a - \phi_c^{(v)}) = \prod_{b=1}^{m_v}e_2(\mu_a - \mu_c).\label{BA2}
\eea

Taking the continuum limit of Eq.s~(\ref{BA1}, \ref{BA2}) and following the standard methods 
we  obtain   
the thermodynamic Bethe ansatz (TBA) equations for the (5+1) model.
The free energy is 
\bea
F/L = -T\sum_{a=0,s,v}\frac{M_a}{2\pi}\int \rd\theta\cosh\theta\ln[1 + \re^{-\epsilon_a(\theta)/T}]
\eea
where the dressed energies $ \epsilon_a$ ($a=0,v,s$) satisfy the nonlinear integral equations 
\begin{widetext}
\begin{subequations}
\bea
&& \epsilon_0(\theta) = M_0\cosh\theta - Ts^{(1/2)}*\ln[1 + \re^{-\epsilon_s(\theta)/T}]\\
&&\epsilon_s(\theta) = M_s\cosh\theta + TY_{ss}*\ln[1 + \re^{-\epsilon_s(\theta)/T}] +  TY_{sv}*\ln[1 + \re^{-\epsilon_v(\theta)/T}] - Ts^{(1/2)}*\ln[1 + \re^{-\epsilon_0(\theta)/T}] - \nonumber\\
&&Ta_n^{(1/2)}*\ln[1 + \re^{-\epsilon_n^{(s)}/T}]\\
&&\epsilon_v(\theta) = M_v\cosh\theta + TY_{vv}*\ln[1 + \re^{-\epsilon_v(\theta)/T}] + TY_{sv}*\ln[1 + \re^{-\epsilon_s(\theta)/T}] - Ta_n*\ln[1 + \re^{-\epsilon_n^{(v)}/T}]
\eea
\end{subequations}
and 
\begin{subequations}
\bea
&& \ln[1 + \re^{\epsilon_n^{(s)}/T}] - A_{n,m}^{(1/2)}*\ln[1 + \re^{-\epsilon_m^{(s)}/T}] + s^{(1/2)}*A_{n,2m}^{(1/2)}*\ln[1 + \re^{-\epsilon_m^{(v)}/T}] =  a_n^{(1/2)}*\ln[1 + \re^{-\epsilon_s/T}]\\
&& \ln[1 + \re^{\epsilon_n^{(v)}/T}] - A_{n,m}*\ln[1 + \re^{-\epsilon_m^{(v)}/T}] + \psi*A_{n,m/2}*\ln[1 + \re^{-\epsilon_m^{(s)}/T}]  = a_n*\ln[1 + \re^{-\epsilon_v/T}]
\eea
\end{subequations}
\end{widetext}
These equations can be solved numerically to get a detailed description of the thermodynamics. For the purposes  of this paper it will be sufficient to study asymptotics of the free energy at large and small temperatures. In the above equations  $*$ stands for the operation of convolution $
f*g(\theta) = \int \rd u f(\theta -u)g(u)$ 
and all kernels are related to the Fourier images as $
f(\theta) = \int \frac{\rd\omega}{2\pi}\tilde f(\omega)\exp\left[-\frac{3\ri\theta\omega}{\pi}\right] $. 
The kernels in Fourier space are
\bea
\label{kernels}
&& \tilde Y_{ss}(\omega)  = - \frac{\re^{-2|\omega|}}{4 \cosh (3\omega/2)\cosh(\omega/2)}\\
&& \tilde Y_{sv}(\omega) = \frac{\re^{|\omega|}}{2\cosh(3\omega/2)}, 
~~ \tilde Y_{vv}(\omega) =
\frac{\cosh(\omega/2)\re^{|\omega}|}{\cosh(3\omega/2)} -1 \nonumber \\ 
&& \tilde a_n(\omega) = \re^{-n|\omega|}, ~~ \tilde s(\omega) =
1/2\cosh\omega, ~~ \tilde \psi(\omega) =
\frac{\cosh(\omega/2)}{\cosh\omega}\nonumber\\ 
&& \tilde A_{nm}(\omega) = \coth|\omega|\left[\re^{-|n-m||\omega|} -
  \re^{-(n+m)|\omega|}\right]\nonumber 
\eea
and we used the notation
\be
\tilde f^{(1/2)}(\omega) = \tilde f(\omega/2).
\ee
At $M_0=0$ $(g_X =0)$ model (\ref{5+1}) decouples into  one
massless Majorana mode and the O(5) GN model. As usual, we expect that  the
massless mode does not appear in  the Bethe ansatz for massive particles. We observe that in this limit  the above TBA  
 equations coincide with the ones derived in
\cite{fendley} for the O(2N+1) GN model for the case  $N=2$. The dressed energies  $\epsilon_0$, $\epsilon_v$, $\epsilon_s$  correspond  to 
$\epsilon_{N,-1}$, $\epsilon_{N-1,0}$, $\epsilon_{N,0}$ of
\cite{fendley}. This is an important selfconsistency check of our solution. 
Please note that while $\epsilon_{N,-1}$ was an
auxiliary particle in the O(5) GN model, in our model ($M_0\neq 0$)
it is a physical particle.

With the TBA equations one can run two
additional checks. First, by calculating the free energy at $T >> M_a$ (for this end one can use the results of \cite{fendley}) one finds the universal asymptotics $F/L = - C_{UV}T^2\pi^2/6$ such that  at 
 $M_0 \neq 0$ one gets $C_{UV} =3$. This is the correct asymptotics for a theory of 6 Majorana fermions. On the other hand  the  $T << M_a$ limit reproduces another asymptotics 
\bea
F/L = - T\sum_a r_a M_a\int \frac{\rd \theta}{2\pi}\cosh\theta \re^{-M_a\cosh\theta/T}
\eea
with $r_s=4, r_v = 5, r_0 =1$ being the correct particle multiplicities.
   
The generalization for the (7+1) model is straightforward. Here  we  have 8
kinks, 7+1 vector particles and 28  tensor particles. The S-matrix
has the same block structure as (\ref{S}), but with 
\bea
\xi_7(\theta) =  \frac{\re^{5\theta} - \ri}{\re^{5\theta} + \ri}
\eea
and an additional block $S_{tt}$ for the tensor particles. The blocks $S_{vv}$ and $S_{tt}$ coincide with the vector and tensor S-matrices of the O(7) GN model and are described in \cite{fendley}. The spinor block $S_{ss}$  is similar in its form to (\ref{spinor}) with $\pi/3$ substituted by $\pi/5$. It does not have poles since the vector particles are not bound states of spinors.

The spin sector at arbitrary filling and the entire model
(\ref{model}) at half filling are quantum  liquids. Away from half
filling the corresponding order parameters and the phase diagram are
described in \cite{azaria,wu}. At moderate forward scattering there
are two phases corresponding to Density Wave or the 
superfluid state of the BCS type. They are distingushed by the sign of
coupling $g_X$ which does not affect  
the excitation spectrum. Both these phases  have power law
correlations and are separated by the Ising type Quantum Phase transition. 
 Generalizing the arguments given in \cite{wu}, we
conclude that for  half filling there are also two phases. Just one of them has  an order parameter local in terms of the fermions: it is $2k_F = \pi$
Charge Density Wave (CDW). It condenses at $g_X < 0$;  at $g_X >
0$ the order parameter is nonlocal corresponding to  a hidden (topological) order.

 The fact that the  O(6)( or O(8)) symmetry is broken down to O(5)$\times\mathbb{Z}_2$ (or O(7)$\times\mathbb{Z}_2$)  will play a role for the correlation functions.   Indeed, the  O(6) and O(8) groups have   
two irreducible spinor representations and O(5) and O(7) have just
one. As it was shown in \cite{KonLud}, in the O(2N) GN model
right and left moving fermions transform according to different spinor 
representations making  it impossible to have a nonzero Green's
function containing  right and left fermions.  However, since  the O(8) symmetric point in model (\ref{5+1}) is
unstable, the number of kinks in the exact solution is the same as
for the O(7) group, which  has just one irreducible
representation. Therefore the Green's function of the right and left
fermions may be non zero. It is logical to assume that it does exist in the dimerized phase corresponding to $g_X < 0$. 
We intend  to discuss this issue  in subsequent publications.   

We are grateful to A. Nersesyan, V. Fateev, F. Smirnov, F. Essler and especially G. Shlyapnikov for discussions and
interest in the work and to
Abdus Salam ICTP for hospitality. This research was supported  by 
European Commission TMR program HPRN-CT-2002-00325 (EUCLID) and Institute for
Strongly Correlated and Complex Systems at BNL (DC) and by US DOE
under contract number DE-AC02 -98 CH 10886 (AMT).

\end{document}